\documentclass[useAMS,usenatbib, a4paper]{mn2e}
\usepackage{natbib}
\usepackage{epsfig}

\usepackage{amssymb}
\usepackage{amsmath}
\usepackage{latexsym}
\usepackage{ifthen}
\usepackage{url}
\usepackage{comment}
\usepackage[T1]{fontenc}
\usepackage{aecompl} 
\usepackage{multirow}

\newcommand\apj{ApJ}%
\newcommand\apjl{ApJ}%
\newcommand\pasp{PASP}%
\newcommand\aap{A\&A}%
\newcommand\apjs{ApJS}%
\newcommand\pasj{PASJ}%
\newcommand\mnras{MNRAS}%
\newcommand\aapr{A\&ARv}

\newcommand{\degrees}{\,^\circ}
\newcommand{\msun}{\,M_{\odot}}
\newcommand{\tsun}{T_{\odot}}

\newcommand{\pb}{P_b}
\newcommand{\pbdot}{\dot{P}_b}
\newcommand{\omdot}{\dot{\omega}}

\newcommand{\pprec}{P_\mathrm{prec}}

\newcommand{\fhour}{^{\mathrm{h}}}
\newcommand{\fmin}{^{\mathrm{m}}}
\newcommand{\fsec}{\mbox{\ensuremath{.\!\!^{\mathrm{s}}}}}
\newcommand{\fdeg}{^{\circ}}
\newcommand{\fdegdecimal}{\mbox{\ensuremath{.\!\!^{\circ}}}}

\newcommand{\rmsresids}{19.3}  
\newcommand{\redchisq}{1.30}  

\newcommand{\incl}{68.0}       
\newcommand{\inclflip}{112.0}  

\newcommand{\pxval}{1.05 \pm 0.55}     
\newcommand{\distval}{0.95 \pm 0.50}   
\newcommand{\pxlkval}{0.082_{-0.027}^{+0.036}}    
\newcommand{\distlkval}{0.73_{-0.24}^{+0.60}}     
\newcommand{\dmdist}{2.5}   
\newcommand{\distpbdotonesig}{0.39}     
\newcommand{\distpbdottwosig}{1.2}     
\newcommand{\distpbdotthreesig}{2.0}   

\newcommand{\pmraval}{-2.42 \pm 0.08}  
\newcommand{\pmdecval}{5.5 \pm 7.0}    
\newcommand{\velra}{-8_{-2}^{+7}}        
\newcommand{\veltot}{20_{-22}^{+27}}    

\newcommand{\pbval}{7.67}      

\newcommand{\precperiod}{496}    
\newcommand{\spercent}{4}      

\newcommand{\pbdotval}{(-2.29 \pm 0.06) \times 10^{-13}}  
\newcommand{\pbdotcorr}{-2.34_{-0.06}^{+0.09} \times 10^{-13}} 
\newcommand{\pbdotgrpred}{(-2.168\pm0.015)\times10^{-13}}  
\newcommand{\pbdotdiffcorr}{(0.18_{-0.06}^{+0.09})\times10^{-13}}
\newcommand{\xpbdot}{(-1.2 \pm 0.5) \times 10^{-14}}

\newcommand{\mpddgr}{1.341 \pm 0.007}       
\newcommand{\mcddgr}{1.230 \pm 0.007}       
\newcommand{\mtotddgr}{2.56999 \pm 0.00006} 

\newcommand{\deltaonesig}{5.9}
\newcommand{\deltatwosig}{34}
\newcommand{\deltathreesig}{66}

\newcommand{\psr}[1]
{\ifthenelse{\equal{#1}{0737}}{PSR~J0737$-$3039}{\ifthenelse{\equal{#1}{1534}}{PSR~B1534+12}{\ifthenelse{\equal{#1}{1913}}{PSR~B1913+16}{\ifthenelse{\equal{#1}{1802}}{PSR~J1802$-$2124}{\ifthenelse{\equal{#1}{1756}}{PSR~J1756$-$2251}{\ifthenelse{\equal{#1}{1906}}{PSR~J1906+0746}{\ifthenelse{\equal{#1}{1141}}{PSR~J1141$-$6545}{\ifthenelse{\equal{#1}{2127}}{PSR~B2127+11C}{\ifthenelse{\equal{#1}{0751}}{PSR~J0751+1807}{\ifthenelse{\equal{#1}{1713}}{PSR~J1713+0747}{\ifthenelse{\equal{#1}{1518}}{PSR~J1518+4904}{\bf ???????}}}}}}}}}}}}

\title[PSR~J1756$-$2251 mass measurements and evolution]{PSR~J1756$-$2251: a pulsar with a low-mass neutron star companion}

\author[Ferdman et al.]{%
R.~D.~Ferdman,$^{1}$\thanks{E-mail: rferdman@physics.mcgill.ca} 
I.~H.~Stairs,$^{2}$
M.~Kramer,$^{3,4}$
G.~H.~Janssen,$^{5,4}$
C.~G.~Bassa,$^{5,4}$
\newauthor
B.~W.~Stappers,$^{4}$
P.~B. Demorest,$^{6}$
I. Cognard,$^{7,8}$
G. Desvignes,$^{3}$
G. Theureau,$^{7,8,9}$
\newauthor
M. Burgay,$^{10}$
A.~G.~Lyne,$^{4}$
R.~N.~Manchester,$^{11}$ and
A. Possenti$^{10}$ \\
$^{1}$Department of Physics, McGill University, 3600 University Street, Montreal, QC, H3A 2T8, Canada\\
$^{2}$Department of Physics and Astronomy, University of British 
  Columbia, 6224 Agricultural Road, Vancouver, BC, V6T 1Z1, Canada\\
$^{3}$Max-Planck-Institut f\"{u}r Radioastronomie, Auf dem H\"{u}gel 69, 53121, Bonn, Germany\\
$^{4}$Jodrell Bank Centre for Astrophysics, University of Manchester, Alan Turing Building, Oxford Road, Manchester M13 9PL, United Kingdom\\
$^{5}$ASTRON, the Netherlands Institute for Radio Astronomy,
Postbus 2, 7990 AA Dwingeloo, the Netherlands\\
$^{6}$National Radio Astronomy Observatory, Green Bank, WV 24944\\
$^{7}$Station de Radioastronomie de Nan\c{c}ay, Observatoire de Paris, 18330 Nan\c{c}ay, France\\
$^{8}$LCP2E/CNRS and University of Orl\'eans, F-45071 Orl\'eans Cedex 2, France\\
$^{9}$GEPI, Observatoire de Paris, CNRS, Universit\'{e} Paris Diderot, 92195 Meudon, France\\
$^{10}$INAF-Osservatorio Astronomico di Cagliari, via della Scienza 5,
09047, Selargius, Italy\\
$^{11}$CSIRO Astronomy and Space Science, Australia Telescope National Facility, Epping NSW 1710, Australia}

\begin{document}

\date{Draft \today; submitted for publication in MNRAS}

\pagerange{\pageref{firstpage}--\pageref{lastpage}} \pubyear{2014}

\maketitle

\label{firstpage}

\begin{abstract}
The pulsar \psr{1756} resides in a relativistic double neutron star (DNS) binary system with a $\pbval$-hr orbit.  We have conducted long-term precision timing on more than 9 years of data acquired from five telescopes, measuring five post-Keplerian parameters.  This has led to several independent tests of general relativity (GR), the most constraining of which shows agreement with the prediction of GR at the $\spercent\%$ level.  
Our measurement of the orbital decay rate disagrees with that predicted by GR, likely due to systematic observational biases.  We have derived the pulsar distance from parallax and orbital decay measurements to be $\distlkval\,$kpc ($68\%$) and $<\distpbdottwosig\,$kpc ($95\%$ upper limit), respectively; these are significantly discrepant from the distance estimated using Galactic electron density models.  We have found the pulsar mass to be $\mpddgr\msun$, and a low neutron star (NS) companion mass of $\mcddgr\msun$.  We also determined an upper limit to the spin-orbit misalignment angle of $\deltatwosig\degrees$ ($95\%$) based on a system geometry fit to long-term profile width measurements.  These and other observed properties have led us to hypothesize an evolution involving a low mass loss, symmetric supernova progenitor to the second-formed NS companion, as is thought to be the case for the double pulsar system \psr{0737}A/B. This would make \psr{1756} the second compact binary system providing concrete evidence for this type of NS formation channel.
\end{abstract}


\section{Introduction}
\label{sec:intro}

\begin{table*}
	\centering
	\caption{Summary of observations and analysis of PSR J1756$-$2251.\label{tab:1756_obs}}
	\begin{tabular}{@{}llcccrcccc@{}}
		\hline
		\multirow{4}{*}{Telescope}  &  
		\multirow{4}{*}{Instrument} & 
		\multirow{2}{*}{Centre}  &  
		\multirow{2}{*}{Total effective} &   
		\multirow{2}{*}{Integration} &
		\multirow{2}{*}{Number} &
		\multirow{2}{*}{Start -- end} &
		\multicolumn{2}{c}{Modifications to} & 
		\multirow{2}{*}{Weighted RMS} \\
		&   
		& 
		\multirow{2}{*}{frequency} &  
		\multirow{2}{*}{bandwidth} &  
		\multirow{2}{*}{time} &
		\multirow{2}{*}{of TOAs} &
		\multirow{2}{*}{dates} &
		\multicolumn{2}{c}{TOA error} & 
		\multirow{2}{*}{of residuals} \\
		\cline{8-9}
		&   
		& 
		\multirow{2}{*}{(MHz)} &  
		\multirow{2}{*}{(MHz)} &  
		\multirow{2}{*}{(min)} &
		&
		\multirow{2}{*}{(MJD)} & 
		Add$^a$ & 
		Multiply$^b$ &
		\multirow{2}{*}{($\mu$s)} \\
		&
		&		
	    &  
        &  
	    &
		&
		& 
		($\mu$s)&
		&
		\\
		\hline
		Parkes                   & Filterbank  &  1274/1390  &  288/256   &  $\sim10$  &  333  &  $52826-54299$  & 2.3    & 1.66  & 19.8\\
		GBT                      & GASP        &  1400       &  $64-96$   &  $1-3$     & 5415  &  $53274-54950$  & \ldots & 1.12  & 16.9\\
		Nan\c{c}ay               & BON         &  1398       &  $64-128$  &  2         & 666   &  $53399-55010$  & \ldots & 1.08  & 28.5\\
		\multirow{2}{*}{Lovell}  & DFB         &  1532       &  384       &  5         & 253   &  $55057-55682$  & \ldots & 1.10  & 23.9\\
		                         & ROACH       &  1532       &  400       &  1         & 571   &  $55696-56334$  & \ldots & 1.16  & 32.9\\
		WSRT                     & PuMa2       &  1380       &  160       &  1         & 1505  &  $54155-56337$  & \ldots & 1.09  & 30.0\\
		\hline
	\end{tabular}
	\flushleft
	$^{a}$Amount added in quadrature to TOA uncertainties. This was only done with Parkes telescope data. \\
	$^{b}$Amount by which TOA uncertainties are multiplied.
\end{table*}

Pulsars in double neutron star (DNS) binary systems represent a distinct population, in which the binary pulsar has been ``recycled'' to faster spin periods.  In most scenarios describing the evolution of such systems, pulsar spin-up is achieved via a phase of mass transfer that also increases the angular momentum of the accreting neutron star \citep[e.g.,][]{acrs82}.  The endpoint of this process is a pulsar with rotation periods $P_{\rm spin} \lesssim 50\,$ms.  For detailed overviews of binary evolution, including that of DNS systems, see, e.g., \citet{bv91,pk94,sta04,tv06}, and \citet{ijc+13}. 

Unlike NS-white dwarf (WD) binaries, DNS systems have evolved through a set of evolutionary scenarios in which the system must proceed through two supernovae (SNe) and avoid disruption.  In forming the first NS, the standard scenario involves the more massive primary star evolving off the main sequence (MS) and filling its Roche lobe, donating matter to its companion until it undergoes iron core collapse, resulting in an SN that leaves behind a NS remnant \citep[see, e.g.,][]{tv06}.  An alternative channel begins with two massive stars of nearly equal mass, in an orbit that is wide enough for the primary to be able to form a CO core, and where the secondary evolves off the main sequence before the primary undergoes a SN.  In this ``double-He core'' scenario \citep{bro95}, the resulting mass transfer rate is very high, causing inspiral and forming a large common envelope (CE) that is promptly ejected, after which the primary then undergoes SN to leave behind a NS.

Our current understanding of the formation of the second NS is divided into two general categories. In the first of these, unstable mass transfer occurs once the massive secondary overfills its Roche lobe.  As with a class of NS-WD referred to as intermediate-mass binary pulsars \citep[IMBPs;][]{cnst96,eb01a} evolution discussed above, a CE is formed, the NS spirals inward, and the envelope is subsequently ejected, in the process avoiding hypercritical accretion onto the NS that would otherwise result in black hole formation \citep[e.g.,][]{fsk+10,tlk12}.  In the DNS progenitor case, the He core that remains may transfer more matter and angular momentum onto the NS as it continues to evolve, until it undergoes a traditional, asymmetric iron-core collapse SN (ICCS). Here, a significant amount of matter is ejected from the system, which is also given a substantial natal kick, resulting in a DNS with an increased eccentricity and high space velocity, and where the normal to the orbital plane is reoriented away from the direction of the spin axis of the first-formed NS.

In contrast, the second category involves \emph{symmetric} SNe events, which proceed on a sufficiently fast timescale, so as to avoid the formation of instabilities that result in the asymmetric explosions described above \citep[][]{plp+04, tlm+13}.  In these scenarios there is also very little mass loss and a weak natal kick to the system.  The resulting eccentricity of the orbit would also generally be lower compared to the ICCS events discussed above. The spin axis of the first-formed NS should therefore also retain its near-alignment with the total angular momentum of the system (well-approximated by the orbital angular momentum) after this low-kick event, since the two are expected to have aligned during the accretion processes after the first SN \citep[e.g.,][]{plp+04,vdh04,pdl+05}.  

Candidates for a symmetric event include electron capture SNe (ECS), so called because the O-Ne-Mg core density of the secondary surpasses a critical limit that allows electrons to capture onto $^{24}$Mg.  Collapse ensues since the electron degeneracy pressure---and Chandrasekhar mass---undergo a rapid decline \citep{mnys80,nom84,pdl+05}.  Another scenario involves Type Ic SNe that occur via core collapse of an ultra-stripped He star, brought about by sufficient mass transfer onto a NS.  Such a process is postulated to result in a sudden and exceptionally faint core-collapse SN, ejecting very little mass from the system \citep{tlm+13}.  One of these channels may explain observations of the double pulsar, \psr{0737}A/B, where the second-formed NS in that system is thought to be the remnant of a symmetric SN \citep{fsk+13,tlm+13}. This is evidenced by its low mass, small orbital eccentricity, low system tangential space velocity, and relative alignment of the pulsar spin axis and orbital angular momentum.

\psr{1756} is in a double neutron star binary system, and was discovered in the Parkes Multibeam Survey \citep{mlc+01,fkl+05}.  Initial timing of this pulsar showed it to have a similar orbital period to the binary pulsar \psr{1913} \citep{ht75a} of $\sim 8$ hours.   However, it was also found to be more spun-up ($P_{\mathrm{spin}}=28.4\,$ms), in a somewhat less eccentric orbit ($e\sim 0.18$), with a companion neutron star apparently having a relatively low mass $m_c=1.18^{+0.02}_{-0.03}\msun$ \citep[][]{fkl+05}.  This showed it to have more characteristics in common with \psr{0737}A, the recycled pulsar in the double pulsar system \citep{bdp+03,lbk+04,ksm+06}.  As discussed in \citet{fsk+13}, it can be argued that the B pulsar in that system had a low-mass progenitor ($< 2\msun$) that underwent a symmetric SN, which would also explain the small transverse velocity observed in the \psr{0737}A/B system \citep{ps05,wkf+06,std+06}.  The resemblance of \psr{1756} to the double pulsar in its orbital eccentricity and low mass companion neutron star thus presents a new opportunity to investigate this channel of DNS evolution for this system as well.\citep[e.g.,][]{vdh04,wwk10}.  

We have extended the existing observational data of \psr{1756} to gain more significant constraints on the system parameters through precision timing.  We also used this new data set to perform an analysis of the pulse shape evolution, in order to study the effects of geodetic precession on the observed pulse profile.  This has helped to constrain the pulsar's spin and orbital geometry, providing further clues as to how this system formed and evolved.  In \S\ref{sec:1756_obs} we describe our observations. In \S\ref{sec:timing} we describe our timing analysis, as well as distance and mass measurements.  We discuss tests of GR with this pulsar in \S\ref{sec:gr_tests}, including correcting our measurement of orbital decay for kinematic biases.  In \S\ref{sec:geodetic} we describe our determination of the geometry of the \psr{1756} system, and we discuss the implications of our findings on its evolution in \S\ref{sec:evol}.  Finally, we summarize our work and provide concluding remarks in \S\ref{sec:conclusions}.

\section{Observations}
\label{sec:1756_obs}

Our data set combines observations from five telescopes.  In what follows, we describe the data acquisition and instruments used.  A summary of the observations is given in Table~\ref{tab:1756_obs}.

\subsection{Parkes}
Observations at the Parkes telescope were performed at regular intervals for \psr{1756} since its initial discovery in the Parkes Multibeam Survey \citep{mlc+01}, and we use data until 2007 July 18 (MJD 54299) for this analysis.  The initial search observations are not included in the timing analysis performed here, but we do incorporate Parkes data used in the initial timing study of this pulsar \citep{fkl+05}.  Data were taken at 1374 MHz center frequency over 288 MHz bandwidth divided into a filterbank of 3-MHz channels, and at 1390 MHz over 256 MHz bandwidth divided into 0.5-MHz channels.  The data from each channel were detected and the two polarizations summed in hardware before 1-bit digitization every 250 and 80--250\,$\mu$s, respectively.  The data were recorded to tape and subsequently folded offline in subintegrations of typically 10 min.  Further details of the Parkes observations can be found in \citet{mlc+01} and \citet{fkl+05}.

\subsection{Green Bank}
Observations at the GBT were performed with the Green Bank Astronomical Signal Processor (GASP) pulsar backend, at a center frequency of 1400 MHz, and were generally taken over $16\times4$\,MHz frequency channels until 2006 January, at which point we began to include, when available, computing nodes from the Caltech-Green Bank-Swinburne Recorder 2 (CGSR2) pulsar backend.  This extra processing capability allowed us to increase the observing bandwidth to incorporate 24 channels.  The data were coherently dedispersed \citep{hr75} in software before detection. Finally, the data stream was folded using the current best ephemeris for the pulsar every 180 seconds, until 2006 August (MJD 53967), after which time we began folding the incoming data into 60-second integrations.  This was done in order to minimize the amount of pulse phase drifting, while still maintaining adequate signal-to-noise ratio in each pulse profile.  The data were flux-calibrated in each polarization using the signal from a noise diode source that is injected at the receiver, and was done in software, using the \texttt{ASPFitsReader} pulsar data analysis software package \citep{fer08}. The calibrated data were then summed together across all frequency channels to give the total power signal at each subintegration.  Each observing session lasted approximately 8 hours, in order to fully sample the orbit of \psr{1756}.  
\begin{figure}
	\begin{center}
		\includegraphics[width=0.5\textwidth]{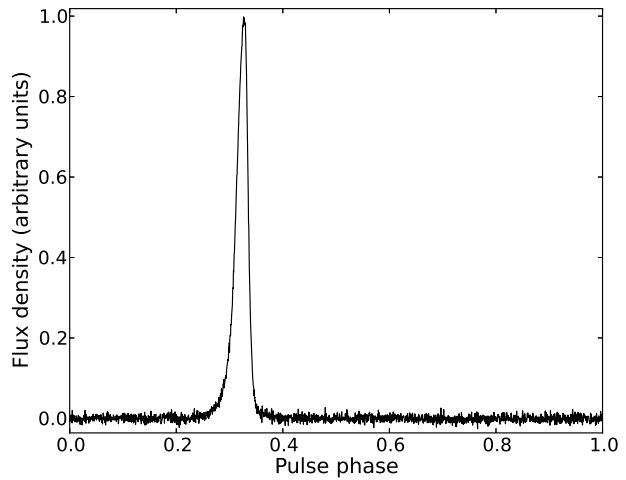}
		\caption{Standard template profile for PSR~J1756$-$2251, constructed from data taken with the GBT using the GASP pulsar backend.\label{fig:1756_std_prof}}
	\end{center}
\end{figure}

\begin{figure*}
	\includegraphics[width=\textwidth]{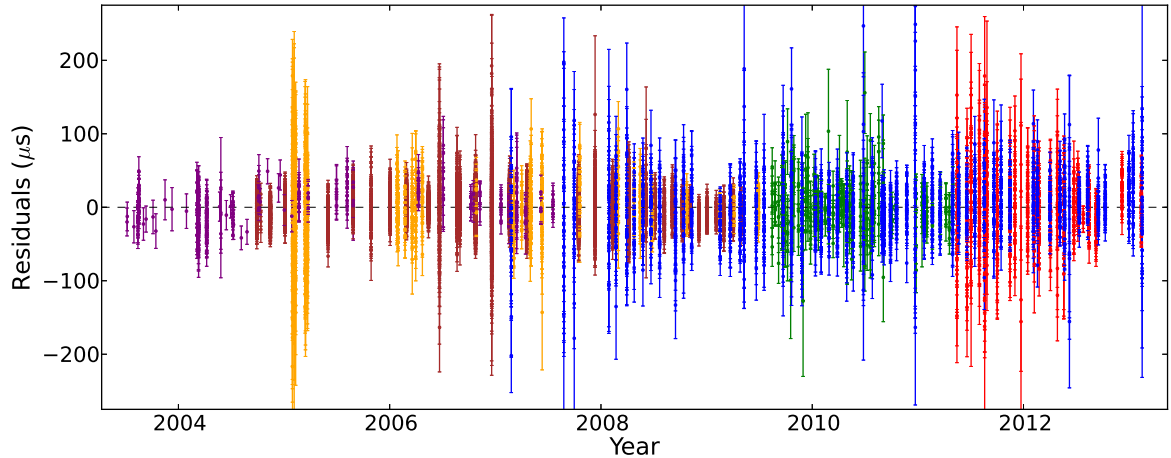}
	\caption{Timing residuals for PSR~J1756$-$2251, after including best-fit parameters in the timing model. Residuals from each instrument used are represented by different colors as follows: Parkes filterbank--purple; GBT/GASP--dark red; Nan\c{c}ay/BON--orange; Westerbork/PuMa2--blue; Jodrell Bank (Lovell)/DFB--green; and Jodrell Bank/ROACH--light red.\label{fig:1756_resids}}
\end{figure*}

\subsection{Nan\c{c}ay}
We included observations of \psr{1756} taken by the Nan\c{c}ay radio telescope in France, with a 94-m circular-dish equivalent diameter.  These data were recorded with the Berkeley-Orl\'{e}ans-Nan\c{c}ay (BON) pulsar backend, a real-time coherent dedispersion instrument, similar to the GASP system at the GBT described above. The data originally consisted of $16 \times 4$-MHz channels, increased to 32 frequency channels as of 2008 July 25, and centered at 1398\,MHz in both cases.  As with the GASP backend, the signal was dedispersed, then detected and folded at the pulse period.  Flux calibration was not yet available for these data; instead, we normalized each hand of polarization by the baseline RMS signal, after which we formed total-power pulse profiles.  Nan\c{c}ay is a meridian-style telescope that observes most sources for approximately 1 hour per day, over which time the telescope gain does not undergo significant change.  We observed \psr{1756} with Nan\c{c}ay at 29 epochs spanning 4.4 years.  The profile data were integrated across the observing bandwidth with a typical integration time of 2 min.  

\subsection{Westerbork}
\psr{1756} was also observed with the Westerbork Synthesis Radio Telescope (WSRT) in The Netherlands once per month, using the Pulsar Machine 2 \citep[PuMa2;][]{ksv08} pulsar backend. Each observation was taken using the full 160-MHz bandwidth that is available, at a center frequency of 1380 MHz, and typically lasting 25 minutes. After each observation, the data were coherently dedispersed \citep[using 64 channels for every 20-MHz band, using the freely available \texttt{dspsr} software;][]{vb11}  and folded with the best available ephemeris for the pulsar using the \texttt{PSRCHIVE} analysis software package \citep{vdo12}. We stored the data as 60-second sub-integrations, divided into 64 frequency channels to allow for realignment in phase of the resulting profiles once an improved ephemeris for the pulsar became available. 

\subsection{Jodrell Bank}
The 76-m Lovell telescope at Jodrell Bank observatory in the United Kingdom observes \psr{1756} at a center frequency of 1532\,MHz and with a monthly cadence. Observations started in early 2009 using an Australia Telescope National Facility digital filterbank (DFB), which performs real-time folding and incoherent dedispersion of two orthogonal polarizations, over a 384-MHz bandwidth using 0.5-MHz channels and 10-s integrations.  Since April 2011 the observations are also processed in parallel with the ROACH backend \citep{kar11}, which uses a ROACH board to sample the two orthogonal polarizations at the Nyquist rate and digitizes them as 8-bit numbers over a bandwidth of 512\,MHz. A 32-channel polyphase filter splits the band in 16-MHz subbands. A high performance computer cluster uses the \texttt{DSPSR} \citep{vb11} software to coherently dedisperse and fold each subband in real time. The useable bandwidth recorded with this instrument is 400\,MHz, split into 0.25-MHz channels and 10-s integrations. RFI in both the DFB and ROACH is removed through automatic scripts and manual inspection.  Furthermore, the spectral kurtosis method for real-time RFI removal \citep{ngl+07} as implemented in \texttt{DSPSR} has been applied to data obtained with the ROACH backend after 2011 November.

\section{Timing Analysis}
\label{sec:timing}

In order to calculate pulse times of arrival (TOAs) for \psr{1756}, we first constructed representative standard template pulse profiles for each telescope and observing setup used, by averaging the data from all scans that did not show RFI contamination or other unusual features.  Based on our long-term profile analysis as described in \S\ref{sec:1756_geom}, there was no concern regarding profile evolution when constructing a template profile in this manner.  The exceptions to this method were made in the cases of Parkes telescope data, for which a template was created out of one high signal-to-noise day of observation, and PuMa2 data taken at the WSRT, for which a modeled noise-free template profile was constructed from the data.  For example, we show the GASP data-derived template profile is shown in Figure~\ref{fig:1756_std_prof}.

Pulse TOAs were then calculated by first cross-correlating each integrated pulse profile in the frequency domain with its corresponding telescope-specific template profile \citep{tay92}.  The time stamp for each integrated profile was then shifted by a time offset corresponding to the phase shift calculated by this cross-correlation.  The uncertainties on the shifts were adopted as the TOA uncertainties.  In total, we measured 8743 individual pulse TOAs; a breakdown of the number of TOAs used from each contributing telescope and backend instrument is found in Table~\ref{tab:1756_obs}.

We then fit a model ephemeris to the topocentric TOAs, which represent the mid-point arrival time for each integration.  This was done using the \texttt{tempo2} software package \citep{hem06,ehm06}, which compares the calculated TOAs to those predicted via an ephemeris that models the various parameters that describe the properties of the pulsar system, and the effect they have on the pulse arrival times.  This includes the rotation frequency and frequency derivatives, as well as delays incurred by the incoming signal due to the free electron content in the interstellar medium, represented by the so-called dispersion measure (DM). We obtained a best-fit value for DM by subdividing the GASP-derived pulse profile data into separate frequency ranges, obtaining TOAs for integrated profiles within each of six frequency bins (1348, 1364, 1389, 1396, 1412, and 1428 MHz).  We performed a timing fit to this data set, and arrived at a best-fit value for DM ($121.198\pm0.005$\,pc\,cm$^{-3}$). This value was held fixed for the subsequent timing analysis performed on the entire frequency-added data set.  Fitting for a change in DM over time did not yield a significant measurement.
 
The model also takes into account the effects of the Earth's motion on the measured pulse TOAs using the the JPL DE421 Solar System ephemeris \citep{sta98b}.  Differences in instrumentation and reference template profiles between observatories caused relative overall offsets in measured pulse arrival times, which were also included as parameters in the fit.  Clock corrections between each observatory and terrestrial time (TT) were obtained using data from the Global Positioning System (GPS) satellites and offsets provided by the Bureau International des Poids et Mesures\footnote{BIPM; \url{http://www.bipm.org/}}.  In the case of Nan\c{c}ay data, recorded times were directly derived from GPS, and thus no additional observatory clock corrections were needed.

\subsection{Binary parameters}
Along with the pulsar spin evolution, DM, and Solar-System effects, delays on the pulse arrival times due to orbital motion were taken into account using the Damour-Deruelle (\textsc{DD}) timing model \citep{dd85,dd86} implemented within the \texttt{tempo2} software.  Here, the usual orbital parameters are modeled: orbital period $P_b$ and eccentricity $e$, longitude of periastron $\omega$, epoch of periastron passage $T_0$, and projected semimajor axis $x \equiv a\sin{i}$.  In addition, this model parametrizes the perturbations to the standard Keplerian description due to relativistic effects, giving the so-called ``post-Keplerian'' (PK) orbital parameters in a theory-independent manner. For a given theory of gravity, the PK parameters are related to the masses of the binary system components and the standard orbital parameters.  In the context of GR, the PK parameters used in our fit are \citep[e.g.,][]{dd86,tw89,dt92}: the rate of periastron advance $\omdot$; a combined time-dilation and gravitational-redshift parameter $\gamma$; the rate of orbital decay $\pbdot$; and the Shapiro delay ``shape'' and ``range'' parameters $r$ and $s$, respectively.

After obtaining a best-fit set of parameters, we reprocessed the data, shifting each integrated profile by the difference in phase between the original profile and that predicted by the updated ephemeris.  This resulted in better-aligned pulse profiles, from which we reconstructed the standard reference profiles; these were then used to then re-perform the timing analysis.
\begin{figure}
	\includegraphics[width=0.5\textwidth]{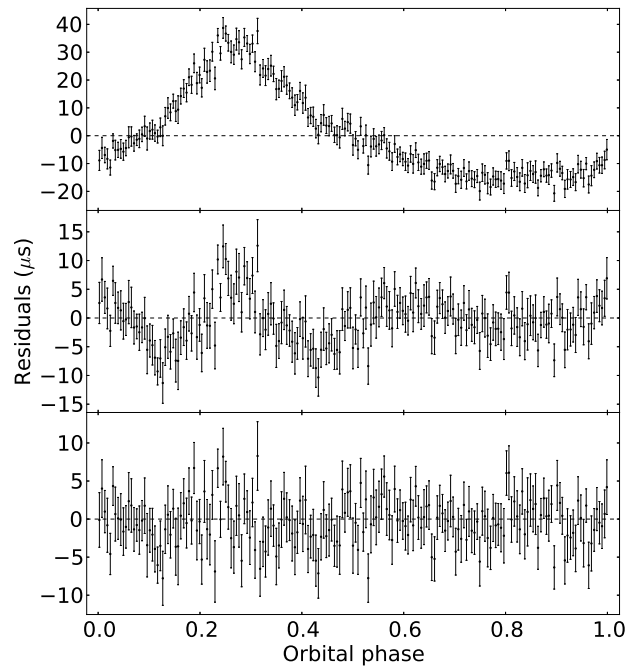}
	\caption{Timing residuals for PSR~J1756$-$2251 as a function orbital phase.  We only show residual derived from GASP pulsar backend data, using the GBT telescope, due to the high data quality and excellent orbital coverage.  Residuals have been averaged into bins of 0.005 in phase, or 2.3 minutes. Top: Shapiro delay $r$ and $s$ parameters were left out of the fit, with all other parameters fixed at their best-fit values, showing the full effect of Shapiro delay on the timing residuals. Middle: $r$ and $s$ were again excluded, but here the remaining parameters were allowed to vary in the fit. Some, but not all, of the Shapiro delay signal has been absorbed, though its effect is still evident.  Bottom: all parameters, including Shapiro delay $r$ and $s$, are included in the timing model fit. \label{fig:1756_shapiro}}
\end{figure}

Figure~\ref{fig:1756_resids} shows the final post-fit timing residuals from all telescope data plotted
over time. These are Gaussian-distributed, as expected for a good fit to the data. However, the TOA uncertainties were in general slightly underestimated, as reflected by an overall $\chi^2$ per degree of freedom $\nu$ that was greater than 1 ($\chi^2/\nu = \redchisq$).  This is likely due to several contributing factors, such as unmitigated RFI, coarse quantization of the analog signal (in the case of Parkes data), and improper characterization of the non-orthogonality of polarization feeds at the telescope front-ends.  To account for these effects, we have scaled, and in the case of the Parkes-derived TOAs, added an amount in quadrature to, the nominal TOA uncertainties obtained from each instrument by an amount that results in $\chi^2/\nu \approx 1$ for each data set.  With the exception of Parkes TOAs, which only contributes to $3.8\%$ of the data set by weight, the scaling factors applied to the data from all telescopes were small ($\lesssim 1.16$), indicating that the calculated uncertainties were reasonably well-understood before this adjustment was made (see Table~\ref{tab:1756_obs} for summary of TOA uncertainty modifications).  We thus directly quote the output parameters and $1\sigma$ uncertainties produced by the \texttt{tempo2} software, which we list in Table~\ref{tab:1756_params}.  The final combined weighted RMS of the timing residuals was $\rmsresids\,\mu{\rm s}$.  
\begin{table*}
	\centering
	\caption{Parameters for PSR~J1756$-$2251.  Unless otherwise stated, observed quantities were measured using the Damour-Deruelle (\textsc{DD}) timing model \citep{dd85,dd86} in \textsc{TEMPO2}.  Figures in parentheses represent the nominal $1\sigma$ uncertainties in the least-significant digits quoted.  Two numbers that are comma-separated indicate the lower and upper uncertainties, respectively. Upper limits are quoted at the $2\sigma$ level, except where noted.  \label{tab:1756_params}}
	\begin{tabular}{lc}
		\hline
\multicolumn{2}{c}{Fit and data-set} \\
\hline
Data span (yr)\dotfill & 9.6 \\
Date range (MJD)\dotfill & $52826.6 - 56337.2$ \\ 
Number of TOAs\dotfill & 8743 \\
RMS timing residual ($\mu s$)\dotfill & 19.3 \\
\hline
\multicolumn{2}{c}{Observed quantities} \\ 
\hline
Right ascension, $\alpha$\dotfill &  $17\fhour56\fmin46\fsec633812(15)$ \\ 
Declination, $\delta$\dotfill & $-22\fdeg51\farcm59\farcs35(2)$ \\ 
Rotation frequency, $\nu$ (s$^{-1}$)\dotfill & 35.1350727145469(6) \\ 
First derivative of rotation frequency, $\dot{\nu}$ (s$^{-2}$)\dotfill & $-1.256079(3) \times 10^{-15}$ \\ 
Reference timing epoch (MJD)\dotfill & 53563 \\ 
Dispersion measure, DM (cm$^{-3}$pc)\dotfill & 121.196(5) \\ 
Parallax (observed), $\varpi$ (mas)\dotfill & 1.05(55)\\ 
Proper motion in right ascension, $\mu_{\alpha}$ (mas\,yr$^{-1}$)\dotfill & $-2.42(8)$ \\ 
Proper motion in declination, $\mu_{\delta}$ (mas\,yr$^{-1}$)\dotfill & $< 20$ \\ 
Orbital period, $P_b$ (d)\dotfill & 0.31963390143(3) \\ 
Orbital eccentricity, $e$\dotfill & 0.1805694(2) \\ 
Projected semi-major axis of orbit, $x \equiv a\sin{i}$ (lt-s)\dotfill & 2.756457(9) \\ 
Longitude of periastron, $\omega_0$ ($\degrees$)\dotfill & 327.8245(3) \\ 
Epoch of periastron, $T_0$ (MJD)\dotfill & 53562.7809359(2) \\ 
Periastron advance, $\dot{\omega}$ ($\degrees\,$yr$^{-1}$)\dotfill & 2.58240(4) \\ 
Time dilation/gravitational redshift parameter, $\gamma$ (ms)\dotfill & 0.001148(9) \\ 
First derivative of orbital period (observed), $\pbdot^\mathrm{obs}$\dotfill & $-2.29(5) \times 10^{-13}$ \\ 
Difference between corrected and GR-derived orbital period derivatives$^{a}$, $\Delta\pbdot^\mathrm{GR, fit}$\dotfill & $-1.2(5) \times 10^{-14}$ \\
Shapiro delay $r$ parameter ($\msun$)\dotfill & 1.6(6) \\ 
Shapiro delay $s$ parameter = sine of inclination angle, $\sin{i}$\dotfill & 0.93(4) \\ 
\hline
\multicolumn{2}{c}{Derived Quantities} \\
\hline
Galactic longitude, $\ell$\dotfill & $6\fdegdecimal498658(5)$ \\
Galactic latitude, $b$\dotfill & $0\fdegdecimal948010(3)$ \\
Rotation period, $P$ (ms)\dotfill & 28.4615890259983(5) \\
First derivative of rotation period, $\dot{P}$\dotfill &  $1.017502(3) \times 10^{-18}$\\
Characteristic age, $\tau_c$ (Myr)\dotfill & 443.5 \\
Surface magnetic field strength, $B_s$ (G)\dotfill & $5.45 \times 10^9$ \\
Inclination of orbit, $i$ ($\degrees$)\dotfill & 68.0(5,6) or 112.0(6,5)\\
Mass function$^{b}$, $f$ ($\msun$)\dotfill & 0.220109(9) \\
Pulsar mass$^{b}$, $m_p$ ($\msun$)\dotfill &  1.341(7)\\
Companion mass$^{b}$, $m_c$ ($\msun$)\dotfill &  1.230(7)\\
Total system mass$^{b}$, $M$ ($\msun$)\dotfill & 2.56999(6)\\
First derivative of orbital period (kinematically corrected), $\pbdot^\mathrm{intr}$\dotfill & $-2.34(6,9) \times 10^{-13}$ \\ 
Difference between corrected and GR-derived orbital period derivatives$^{d}$, $\Delta\pbdot^\mathrm{GR, intr}$\dotfill & $1.8(6,9) \times 10^{-14}$ \\
Total proper motion, $\mu$ (mas\,yr$^{-1}$)\dotfill & $<19$ \\
Tangential space velocity, $v_t$ (km\,s$^{-1}$)\dotfill & $<68$ \\
Parallax (Lutz-Kelker bias corrected), $\varpi_\mathrm{corr}$ (mas)\dotfill & $0.082(27,36)$\\
Distance to pulsar (raw parallax), $d$ (kpc)\dotfill &  0.95(50)\\
Distance to pulsar (Lutz-Kelker bias corrected), $d_\mathrm{corr}$ (kpc)\dotfill & 0.73(24,60)\\
Distance to pulsar (orbital decay), $d_{\pbdot}$ (kpc)\dotfill & $<1.2$\\
Precession period, $P_\mathrm{prec}$ (yr)\dotfill & 496\\
\hline
\multicolumn{2}{c}{Derived quantities---geometry fit} \\
\hline
Spin/magnetic axis angle, $\alpha$ ($\degrees$)\dotfill & $109(24,16) (i=68\degrees)$ or $74(16,24) (i=112\degrees)$ \\
Spin/total system angular momentum misalignment angle, $\delta$ ($\degrees$)\dotfill & $<5.9\ (68\%)$ , $<34\ (95\%)$, $<66\ (99\%)$\\
\hline
\end{tabular}
\flushleft
$^{a}$Measured with the Damour-Deruelle General Relativity (\textsc{DDGR}) timing model \citep{dd86,tw89} in \textsc{TEMPO2}, which assumes general relativity to be the correct theory of gravity, via the XPBDOT parameter.\\
$^{b}$Measured using a likelihood grid, using the \textsc{DDGR} model.\\
$^{c}$Determined from masses derived with the GR formulation of $\omdot$ and $\gamma$ measurements, via the theory-independent Damour-Deruelle timing model \citep[DD;][]{dd86,tw89}.\\
$^{d}$Comparison with $\pbdot^\mathrm{GR}$.
\end{table*}

The Shapiro delay parameters $r$ and $s$ describe the extent to which the incoming pulsar signal undergoes extra delay as it traverses the gravitational potential of its companion, as the pulsar passes through superior conjunction relative to our line of sight.  Figure~\ref{fig:1756_shapiro} shows post-fit timing residuals for GBT-derived data as a function of orbital phase, produced after three iterations of model fitting.  One can see that if we perform a fit that includes the Keplerian orbital parameters, but not $r$ or $s$, that the Shapiro delay signal is not entirely absorbed. This confirms its effect on the TOAs, and the validity of including these parameters in the timing model fit.

\subsection{Tests of general relativity}
\label{sec:gr_tests}

\begin{figure*}
  \begin{center}
    \includegraphics[width=1.0\textwidth]{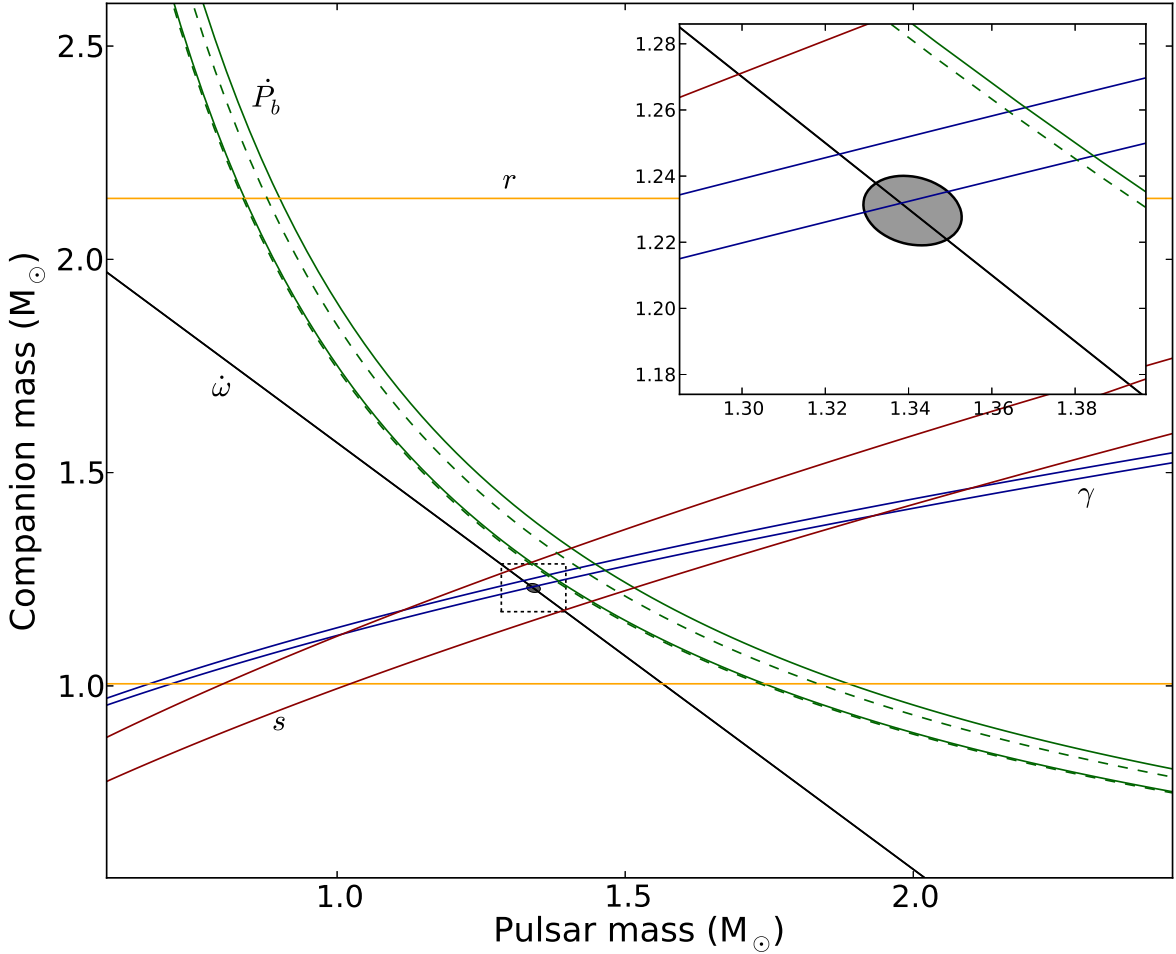}
    \caption{Pulsar mass/companion mass diagram for PSR~J1756$-$2251.  Shown are the $1\sigma$ general-relativistic mass constraints for the five post-Keplerian parameters, which we have measured with significance:  advance of periastron ($\dot{\omega}$), the gravitational redshift/time dilation parameter ($\gamma$), the Shapiro delay $r$ and $s$ parameters, and the orbital period decay rate ($\pbdot$).  We show the latter both before (dashed line; corresponding to $\pbdot^\mathrm{obs}$ in Table~\ref{tab:1756_params}) and after (solid line; corresponding to $\pbdot^\mathrm{intr}$) applying corrections for kinematic biases in its measured value.  The \textsc{DDGR} model-derived component masses of the neutron stars in this system, which assumes GR to be the correct description of gravity, is shown as a filled ellipse, marking the $68\%$ confidence contour.  The inset shows the region close to the \textsc{DDGR} prediction for the system masses (bordered by a dotted rectangle in the main plot).\label{fig:1756_m1m2}}
  \end{center}
\end{figure*}

As reported in Table~\ref{tab:1756_params}, we were able to significantly determine five PK parameters with the \textsc{DD} timing model.  The measurement of any two PK parameters results in a unique determination of the component masses of the system; each additionally measured parameter overdetermines the system, therefore providing an independent GR test.  Figure~\ref{fig:1756_m1m2} plots the GR-derived mass constraints determined from each measured PK parameter.  
The $1\sigma$ constraints from four of these parameters, $\omdot$, $\gamma$, $r$, and $s$, intersect on the diagram.
They also agree within $1\sigma$ with the Damour-Deruelle General Relativity (\textsc{DDGR}) timing model determination of the system masses \citep{dd86, tw89}, which assumes GR to be the correct theory of gravity, as we describe in \S\ref{sec:masses}. The two most precisely measured PK parameters for this system are $\omdot$ and $\gamma$;
from the intersection of the mass constraints of these two quantities, we determined the GR-predicted masses.  We were then able to calculate the predicted values of the remaining PK parameters for GR, allowing us to perform independent tests of GR for each.  The results of these tests are summarized in Table~\ref{tab:1756_gr_tests}.
We find the observed Shapiro delay $r$ and $s$ parameters to agree with the predictions of GR to 4 and 50$\%$, respectively.  
The observed and kinematic bias-corrected orbital decay rates ($\pbdot^\mathrm{obs}$ and $\pbdot^\mathrm{intr}$, respectively; see \S\ref{sec:pbdot} and \S\ref{sec:distance}) disagree with the GR prediction by $2-3\sigma$.  This deviation can also be seen in their derived mass constraints, shown in Figure~\ref{fig:1756_m1m2} (dashed and solid green lines, respectively).  It may be that the GR formulation for quadrupolar gravitational-wave radiation is incorrect, or that GR itself has broken down in the case of this system; while we should not dismiss these notions out of hand, several similar systems have convincingly passed this type of strong-field test.  We thus find it more likely that this discrepancy is due to observational biases, some of which we discuss in \S\ref{sec:pbdot}, and/or other currently unknown systematic effects that are not taken into account with standard timing analysis techniques.

\subsection{Neutron star masses}
\label{sec:masses}
In order to derive the masses of the pulsar and its companion neutron star, we re-performed the above timing analysis, this time employing the \textsc{DDGR} timing model, which directly interprets the values of the measured PK parameters in terms of the companion and total system masses ($m_c$ and $M\equiv m_p+m_c$, respectively) in terms of GR, along with the usual Keplerian orbital parameters.  
In order to account for possible systematic bias in the measured value of $\pbdot$, we fit for an additional parameter which represents the deviation of $\pbdot$ from that predicted by GR.
We performed a maximum likelihood analysis in order to find the best model fit to the masses, in which we derived a joint probability distribution using $\chi^2$ values found from timing fits performed over a fine, evenly distributed grid of $m_c$ and $M$ values.  At each grid point, we hold fixed the corresponding ($m_c, M$) values in the fit, allowing all other model parameters to vary.  From this, we obtain marginalized probability distribution functions (PDFs) for $m_c$ and $M$.  We then calculate a PDF for the pulsar mass from a histogram of derived $m_p=M-m_c$ values, weighted by the normalized likelihood at each ($m_c, M$) grid point.  
From this analysis, we derive a pulsar mass $m_p = \mpddgr\,\msun$, companion mass $m_c = \mcddgr\,\msun$, and total system mass $M = \mtotddgr\,\msun$.  These values overlap within $1\sigma$ of those found from the intersection of the GR-derived mass constraints as discussed in \S\ref{sec:gr_tests}.  Our value for the companion mass is somewhat higher than that determined by \citet{fkl+05}, although they are in agreement at the $2\sigma$ level.  Our analysis represents a larger time span of data, and much better orbital-phase sampling using GASP data; we are therefore confident that this result is more robust than the previously reported mass measurement.

Our derived value for the companion mass makes it one of the lowest-mass NSs known, along with \psr{0737}B, \psr{1518}, and \psr{1802} \citep{ksm+06,jsk+08,fsk+10}.  This may have important implications for the formation history of the system.  In particular, the similarity of \psr{1756} in mass and other properties to the \psr{0737}A/B system may imply that these two DNS systems have proceeded through comparable evolutionary pasts.  We discuss this further in \S\ref{sec:geodetic} and \S\ref{sec:evol}.

\begin{table}
	\centering
	\caption{Independent tests of GR with PSR J1756$-$2251. Observed post-Keplerian (PK) parameters were measured via the DD timing model fit, and are also listed in Table~\ref{tab:1756_params}.  The expected values of each quantity from GR is found by calculating the masses corresponding to the intersection of periastron advance rate $\omdot$ and time dilation/gravitational redshift parameter $\gamma$.  Figures in parentheses represent the nominal $1\sigma$ uncertainties in the least-significant digits quoted.\label{tab:1756_gr_tests}}
	\begin{tabular}{@{}lccc@{}}
		\hline
		\multirow{2}{*}{PK parameter}  &  
		Observed & 
		GR-predicted  &  
		Ratio of observed \\
		&
		value &
		value &
		to expected values \\
		\hline
		$\pbdot^\mathrm{obs}$ ($\times 10^{-13}$)     & $-2.29(5)$   &  \multirow{2}{*}{$-2.168(15)$}   &  1.06(3)   \\
		$\pbdot^\mathrm{intr}$ ($\times 10^{-13}$)    & $-2.34(6,9)$ &                                  &  1.08(3)   \\
		$r$ ($\msun$)                                 & $1.6(6)$     &  1.240(7)                        &  1.3(5)    \\
		$s$                                           & $0.93(4)$    &  0.914(4)                        &  1.01(4)   \\
		\hline
	\end{tabular}
	\flushleft
\end{table}

\subsection{Orbital period decay}
\label{sec:pbdot}

Our timing analysis determines the observed rate of orbital period decay to be $\pbdot^\mathrm{obs} = \pbdotval$.  However, contributions from Galactic acceleration and the proper motion of the pulsar serve to bias the observed value of $\pbdot$ away from its intrinsic value \citep[as well as any change in either NS mass, which we disregard here; see, e.g.,][]{dt91}.  One can easily see the effect of the biased $\pbdot$ in Figure~\ref{fig:1756_m1m2}, where its GR-derived mass constraints are significantly shifted in the positive direction relative to the overlapping region of the other PK parameters.  The influence of these contaminating effects on the observed value of the orbital decay rate can be expressed as follows:
\begin{equation}
	\label{eqn:pbdot_bias}
	\pbdot^\mathrm{obs} = \pbdot^\mathrm{intr} + \pbdot^\mathrm{Gal} + \pbdot^\mathrm{Shk}.
\end{equation}
The first term on the right side of equation~\ref{eqn:pbdot_bias} denotes the intrinsic value of the orbital decay rate for the system, once all kinematic corrections are included (this is not to say that all systematics are accounted for, as we discuss in \S\ref{sec:gr_tests}; nor does it necessarily represent the GR-predicted value, as we will see below). The second term represents the effects due to the Galactic acceleration of the pulsar relative to Earth, and is the combination of components parallel and perpendicular to the Galactic plane:
\begin{equation}
	\label{eqn:pbdot_gal}
	\pbdot^\mathrm{Gal} = \pbdot^\mathrm{Gal, \parallel} + \pbdot^\mathrm{Gal, \perp}.
\end{equation}
The component due to centripetal acceleration parallel to the Galactic plane can be approximated by \citep{dt91,nt95}:
\begin{equation}
	\left(\frac{\pbdot}{\pb}\right)^\mathrm{Gal, \parallel} = -\cos{b}\frac{v_0^2}{cR_0} \left[\cos{\ell} + \frac{\beta}{\sin^2{\ell} + \beta^2}\right],
\end{equation}
where $c$ is the speed of light in a vacuum, $v_0$ and $R_0$ are the Sun's Galactocentric velocity and distance, respectively, $\ell$ and $b$ are the Galactic latitude and longitude, respectively, and $\beta \equiv (d/R_0)\cos{b} - \cos{\ell}$, where $d$ is the Earth-pulsar distance.  The orthogonal component, due to acceleration in the Galactic potential, is given by \citep{dt91}:
\begin{equation}
	\left(\frac{\pbdot}{\pb}\right)^\mathrm{Gal, \perp} = \frac{a_z \sin{b}}{c},
\end{equation}
where $a_z$ is the vertical acceleration component, which depends on the distance of the pulsar from the Galactic plane, as well as the local mass density and disk surface density profile.  Based on the model of the Galactic potential by \citet{kg89}, \citet{nt95} showed that $a_z$ can be expressed as:
\begin{equation}
	\frac{a_z}{c} = -1.09\times10^{-19}\left[\frac{1.25z}{\sqrt{z^2 + 0.0324}} + 0.58z\right],
\end{equation}
where $z \equiv d\sin{b}$.  

Finally, the third term in equation~\ref{eqn:pbdot_bias}, often referred to as the Shklovskii effect \citep{shk70}, is due to the tangential motion of the pulsar relative to our line of sight, causing an apparent positive bias in the value of the decay rate.  It can be calculated from the following:
\begin{equation}
	\label{eqn:pbdot_shk}
	\left(\frac{\pbdot}{\pb}\right)^\mathrm{Shk} = \frac{\mu^2d}{c},
\end{equation}
where $\mu$ is total pulsar proper motion.

We constructed a Monte-Carlo histogram of the intrinsic $\pbdot$ in order to calculate its uncertainty, by choosing random Gaussian-distributed values for the input quantities in equations~\ref{eqn:pbdot_bias} through \ref{eqn:pbdot_shk}, with widths equal to the measured $1\sigma$ uncertainties of those values.  Specifically, the Galactic coordinates $\ell$ and $b$ are found from right ascension and declination, and we take $v_0=240 \pm 8$\,km\,s$^{-1}$ and $R_0=8.34 \pm 0.16$\,kpc \citep{rmb+14}.
For the total proper motion $\mu$, we used our nominally calculated value and uncertainty of $6.0\pm6.4$\,mas\,yr$^{-1}$, even though we quote it as an upper limit in Table~\ref{tab:1756_params}. 
As we will discuss in \S\ref{sec:distance}, there is a non-negligible Lutz-Kelker bias on our parallax and distance measurements, which we calculated and used in our $\pbdot$ correction calculation (we present these in Table~\ref{tab:1756_params} as $\varpi_\mathrm{corr}$ and $d_\mathrm{corr}$, respectively).  We assume that the orbital period $P_b$ is constant, as we observe it to much higher relative precision than we do for the other input quantities.
We then construct a histogram resulting from 65536 iterations, from which we take the median and $68\%$ interval, resulting in $\pbdot^\mathrm{intr} = \pbdotcorr$.  Using this corrected value, we recalculate the GR-derived mass constraints (shown in Figure~\ref{fig:1756_m1m2} as solid green lines).  As mentioned in \S\ref{sec:gr_tests}, this correction increases the uncertainty in $\pbdot$; it also shifts its corresponding GR mass constraints slightly \emph{further away} from the intersection of the other PK parameter constraints,
with which they are inconsistent at the $1\sigma$ level (but marginally consistent at the $2\sigma$ level).
We have explored the possibility that the values measured by \citet{rmb+14} for $v_0$ or $R_0$ are erroneous, by repeating the above analysis with previous measurements of these quantities as input, such as those by \citet{rmz+09} and \citet{hna+12}.  We have found the resulting effect on the output mass constraints to be insignificant in each case.

As discussed in \S\ref{sec:gr_tests}, and shown in Table~\ref{tab:1756_gr_tests}, the GR-predicted value for the orbital decay rate in \psr{1756} is $\pbdot^\mathrm{GR} = \pbdotgrpred$. The difference from the corrected, and thus intrinsic, measurement of $\pbdot$ is therefore:
\begin{equation}
	\Delta\pbdot^\mathrm{GR, intr} = \left|\pbdot^\mathrm{intr} - \pbdot^\mathrm{GR}\right| = \pbdotdiffcorr.
\end{equation}
This observed departure from the GR prediction can be attributed
to the combined effects on the observed orbital decay of this system that are not resulting from kinematic biases.  These may include, for example, a secular change in the gravitational constant $\dot{G}$ \cite{nor90,dt91,lwj+09}, and gravitational dipole radiation on the $\pbdot$ of this system, which is predicted to exist in some alternative scalar-tensor theories of gravity, due to the relative asymmetry in the component masses of this system \citep[e.g.,][]{esp05}.  However, currently unknown systematic observational biases may also contribute to this discrepancy;
as a result, the robustness of any gravity test using the $\pbdot$ we derive from timing this pulsar is limited until we are able to better constrain the systematic effects that influence its measurement. These can include the poorly constrained proper motion and/or distance (however, our work presented in \S\ref{sec:distance} gives us increased confidence in our measurement of the latter), or an incorrect model of the Galactic potential near the pulsar position.  Further observational data will certainly aid in improving the measurements of the astrometric quantities and help to address this issue. This includes long-term Very Long Baseline Interferometry (VLBI) imaging, from which the measured astrometry is less affected by the ecliptic latitude of the observed source than is our timing analysis.  This is important for \psr{1756}, which is very close to the ecliptic plane, with ecliptic latitude $\beta \sim 0.6\degrees$.

\subsection{Parallax and distance measurements}
\label{sec:distance}
With our extended timing baseline, we have been able to measure the parallax of \psr{1756}, which we find to be $\varpi = \pxval\,$mas.  This corresponds to a distance to the pulsar of $d = \distval\,$kpc.  We perform an F-test for inclusion of the parallax into the timing model, and obtain an F-ratio of 0.056; this gives us confidence that the improvement in our fit by incorporating parallax into our model is not likely due to chance.  In contrast, the distance based on the DM of the pulsar, calculated by using the NE2001 Galactic free electron distribution model \citep[][]{cl02}, is approximately $\dmdist\,$kpc\footnote{The typical quoted uncertainty on the NE2001 model distance is $20-30\%$}.  We believe this overestimation of the distance to likely be due to inaccuracies in the modeled electron content in the direction of \psr{1756}. 
 
This discrepancy widens further when correcting for the bias related to our parallax measurement uncertainty, known as Lutz-Kelker bias \citep{lk73}.  \citet{vlm10} have shown that this bias can be calculated for a given pulsar through a Monte-Carlo simulation, assuming a Gaussian parallax measurement uncertainty, and taking into account both the known Galactic pulsar spatial distribution of \citet{lfl+06} and the intrinsic pulsar luminosity function as described by \citet{fk06}.  As the measured fractional uncertainty becomes larger, the corrections to the parallax and distance depend increasingly on the above-cited population and luminosity function models.  As a result, the linear relationship between the model-corrected distance and parallax breaks down, so that one is no longer the simple inverse of the other, as is the case with our observations, where the Galactic population term dominates the parallax correction.  An implementation of this procedure is available online\footnote{\url{http://psrpop.phys.wvu.edu/LKbias/}}, which we have used to calculate this effect on the parallax and distance of \psr{1756}.  Using our timing-derived parallax measurement and a flux at 1400\,MHz of 0.6 mJy \citep{fkl+05}, we find a corrected parallax $\varpi_\mathrm{corr} = \pxlkval$\,mas, and distance to the pulsar $d_\mathrm{corr} = \distlkval\,$kpc.
 
The correct determination of the distance to the pulsar is crucial for reliably correcting for kinematic effects that may contaminate the measurement of the orbital period decay.  We use the Lutz-Kelker bias-corrected distance of \psr{1756} in our effort to do so, as described in \S\ref{sec:pbdot}.  This is also true for space velocity determination, which is important for discussion of the evolution and formation of this system, which we discuss in \S\ref{sec:evol}.  As mentioned earlier, VLBI imaging observations of this system over time would likely produce a more precise distance measurement on a shorter timescale than would long-term timing observations.

As discussed in \S\ref{sec:masses}, the \textsc{DDGR} timing model can be reparametrised to calculated the offset between the GR-predicted value of the orbital decay and the uncorrected measurement of $\pbdot$; doing so, we find a difference $\Delta\pbdot^{\mathrm{GR, fit}} = \xpbdot$.  We can now use this value to set an upper limit to the pulsar distance $d$, by using our $\pbdot$ bias correction equation~\ref{eqn:pbdot_bias} \citep{bb96}, so that
\begin{equation} 
	\label{eqn:pbdot_gr}
	\Delta\pbdot^{\mathrm{GR, fit}} = \pbdot^\mathrm{obs} - \pbdot^\mathrm{intr} = \xpbdot,
\end{equation} 
where we assume that the intrinsic orbital decay of the system is solely due to the effects of GR.  We then inverted equation~\ref{eqn:pbdot_gr} to find the distance to the pulsar $d_{\pbdot}$, and its associate uncertainty via equations~\ref{eqn:pbdot_gal} through \ref{eqn:pbdot_shk}.  This is done using the same Monte-Carlo histogram method as in \S\ref{sec:pbdot} for correcting $\pbdot$. At each iteration, we used Newton's method in order to solve $\Delta\pbdot^{\mathrm{GR, fit}} - (\pbdot^\mathrm{Gal} + \pbdot^\mathrm{Shk}) = 0$ for distance, assuming a random Gaussian distribution for all input variables, including $\Delta\pbdot^{\mathrm{GR, fit}}$, with widths equal to their $1\sigma$ uncertainties (as in \S\ref{sec:pbdot}, we assume the orbital period $P_b$ to be exact since its fractional uncertainty is much smaller relative to the other input values), and build a distribution of all physical (i.e.~positive) output distance values.

We find 68, 95, and 99\% upper limits to the pulsar distance $d_{\pbdot}$ of $\distpbdotonesig$, $\distpbdottwosig$, and $\distpbdotthreesig$\,kpc, respectively.  This is consistent with our parallax-derived value (both bias-corrected and uncorrected) at just over the $1\sigma$ level, and is only consistent with the NE2001 model distance above the $3\sigma$ (based on a $30\%$ uncertainty in the NE2001 value).  Although more observational data will help to better constrain the parallax and distance to the pulsar, this result gives us added confidence in the reliability of our parallax-derived distance measurement, and particularly in its use for correcting orbital decay. It also reiterates the relative inconsistency of the modeled electron density along along the line of sight to \psr{1756}.

\section{Geodetic precession and long-term profile analysis}
\label{sec:geodetic}
According to GR, the spin axis of a pulsar in a binary system will precess about the total angular momentum vector of the system at a rate given by \citep{dr74,bo75}:
\begin{equation}
  \label{eqn:geodetic}
  \Omega_{1}^{\mathrm{spin}} = \left(\frac{2\pi}{P_{\mathrm{b}}}\right)^{5/3} \tsun^{2/3}\frac{m_c(4m_p + 3m_c)}{2(m_p + m_c)^{4/3}} \frac{1}{1-e^2},
\end{equation}
where in this formulation, $m_p$ and $m_c$ are, respectively, the pulsar and companion masses, expressed in solar masses, $e$ is the orbital eccentricity, $P_b$ is the orbital period, and $\tsun = GM_{\odot}/c^3 = 4.925490947\,\mu\textrm{s}$ is the mass of the Sun expressed in units of time.
Our measurement of the system masses, together with equation~\ref{eqn:geodetic}, allows us to calculate the GR-predicted geodetic precession period of the \psr{1756} spin axis to be $\pprec = \precperiod\,$yr, of which our 9-year time baseline of data covers $1.3\%$.  We note that the precession period of this pulsar is longer than most of those in binary systems for which secular effects of geodetic precession on the observed pulse profile have have been observed, such as \psr{0737}B \cite[75\,yr;][]{bkk+08,pmk+10}, \psr{1141} \citep[265\,yr;][]{mks+10}, \psr{1906} \citep[165\,yr;][Desvignes et~al., in prep.]{lsf+06,kas12, vl+14}, \psr{1913} \citep[296\,yr;][]{kra98, wt02}, \psr{2127} \citep[278\,yr;][]{jcj+06,kvf+14}.  However, \psr{1534} \citep{fst14} has $\pprec = 610$\,yr, and time-dependent shape changes are also clearly found in its pulse profile, over a similarly small fraction of its precession period as is spanned by our \psr{1756} data set.
Although our data set represents only a relatively small portion of the total precession period, we might thus still expect some long-term observable changes in the pulse profile.

Our principal motivation for searching for these effects in \psr{1756} is to constrain the spin and orbital geometries of this system, and the insight this can provide into its evolution.  In particular, the spin-orbit misalignment angle $\delta$ can shed light on the formation history of this system.  Specifically, a low spin-orbit misalignment in the pulsar may indicate a low-mass loss, relatively symmetric supernova event having led to the formation of the companion neutron star.  This is thought to be the case for the double pulsar \psr{0737}A/B, to which \psr{1756} has similar masses and orbital eccentricity \citep{ksm+06, fsk+13}.

\subsection{Pulse shape evolution}
\label{sec:pulse_shape}
\begin{figure}
  \begin{center}
    \includegraphics[width=0.5\textwidth]{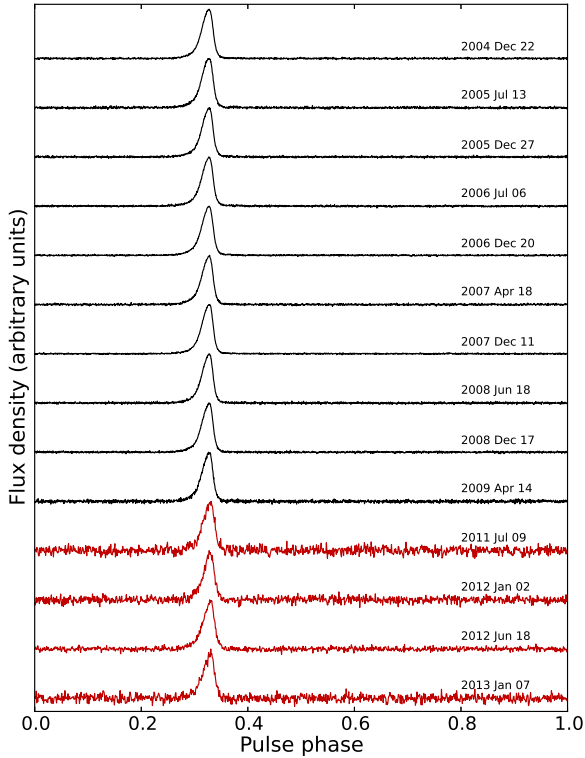}
    \caption{PSR~J1756$-$2251 pulse profiles added over approximately 6-month periods, used to calculate widths for the geometry fit.  Those derived from GBT data, using the GASP backend, are shown in black, and those accumulated from data taken at the Lovell telescope at Jodrell bank using the ROACH backend are plotted in red.\label{fig:prof_evol}}
  \end{center}
\end{figure}

To obtain consistent, high signal-to-noise profiles, we added data from each of the GASP and ROACH backends in groups of 180 days, ensuring that we used the same range of observing frequencies for both backends. The midpoint in time spanned by the data was taken to be its representative date for each added profile. 
We did not expect any pulse shape variation due to scintillation effects, due to the relatively high DM measured for this pulsar, and the relatively low instrumental bandwidth over which these observations were taken. Figure~\ref{fig:prof_evol} shows the resulting pulse profiles.  We saw no obvious long-term changes above the noise level.  Our subsequent width calculations at each epoch follow the bootstrap-style technique described in \citet{fsk+13}.  To summarize, we performed a 32768-iteration, 6th-order polynomial fit to 24 points along each side of the profile, omitting a random choice of 11 data points at every iteration.  We used each fit to interpolate the value of the spin phase at the desired pulse height, then found the difference between the phase values found for each side of the profile, arriving at a pulse width.  We constructed a histogram out of all trial widths, to which we fit a simple Gaussian profile, quoting its mean and width as the median pulse width and corresponding $1\sigma$ uncertainty, respectively.

\begin{figure}
  \begin{center}
    \includegraphics[width=0.5\textwidth]{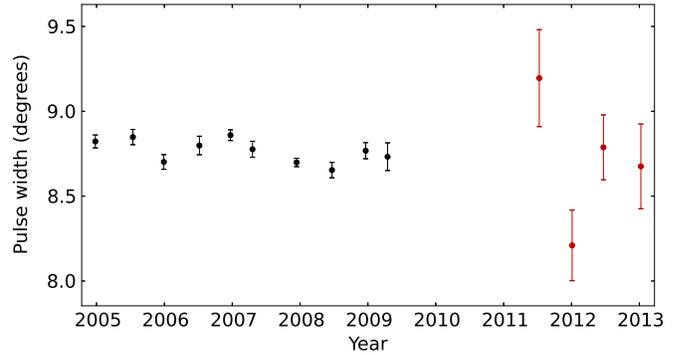}
    \caption{Profile widths at $50\%$ of the peak amplitude for PSR~J1756$-$2251 as a function of time. Black points denote data derived from GBT/GASP backend data, and red points are widths measured from Jodrell Bank data using the Lovell telescope width the ROACH backend.\label{fig:1756_widths}}
  \end{center}
\end{figure}

For each epoch shown in Figure~\ref{fig:prof_evol}, we measured pulse widths at $25\%, 30\%, 35\%, 40\%, 45\%,$ and $50\%$ of the peak pulse height.  Figure~\ref{fig:1756_widths} plots the pulse width measurements  at $50\%$ of the peak pulse height as a function of time.  There is no obvious secular trend, which hints at one of two possibilities:  the misalignment angle $\delta$ between the pulsar spin axis and the total angular momentum of the system is small, or else the pulsar's axis of rotation is currently at a special phase of precession (e.g. $\sim 0\degrees$ or $180\degrees$).  
While we do not necessarily expect the current data set, which represents a small sample of the total precession cycle, to coincide with such a special phase, the latter remains a distinct possibility \citep[as was the case for \psr{1913};][]{kra98, wt02}.

\subsection{Constraints on the geometry of the PSR~J1756$-$2251 system}
\label{sec:1756_geom}
We use the model of \citet{rl06a} to relate the pulse widths to the system geometry as follows:
\begin{equation}
	\label{eqn:phi0}
	\cos{\Phi_0} = \frac{\cos\rho - \cos\zeta\cos\alpha}{\sin\zeta\sin\alpha}.
\end{equation}
Here, $\Phi_0$ is half the pulse width, $\rho$ is the half-opening angle of the part of the emission cone at the pulse height corresponding to the measured width, $\alpha$ is the angular separation between the pulsar spin and magnetic axes, and $\zeta = \zeta(i, \delta, T_1)$ is the angle between the pulsar spin vector and the observer line of sight. $\zeta$ in turn depends on the orbital inclination $i$, the misalignment angle $\delta$ between the spin and total system angular momentum vectors, and $T_1$, the epoch of zero precession phase; the latter is defined via:
\begin{equation}
	\phi_{\mathrm{SO}} = \Omega_1^{\mathrm{spin}}(t - T_1),
\end{equation}
where $\phi_{\mathrm{SO}}(t)$ is the angular precession phase of the spin axis,
and $\Omega_1^{\mathrm{spin}}$ is the angular precession frequency as defined in equation~\ref{eqn:geodetic}.
This method is similar to the pulse profile analysis done by \citet{fsk+13} for the pulse profile analysis of \psr{0737}A; for further details, see \S{5.2} of that paper.  For a full treatment of the geometry involved, refer to \citet{dt92}.
In this analysis, however, the signal to noise of the combined profiles was generally significantly lower than for \psr{0737}A.  For this reason, we used the \citet{rl06a} model to perform a simultaneous fit of pulse widths measured at all pulse heights mentioned in \S\ref{sec:pulse_shape}, to arrive at single $\alpha$ and $\delta$ values.  This is in contrast to taking the average of the $\alpha$ and $\delta$ values derived from separately fitting each set of measured pulse widths, as was done in \citet{fsk+13}.  As in that study, we allow the set of $\rho$ values to vary at each point of a three-dimensional grid of $\alpha$, $\delta$, and $T_1$ values.  We derived at a joint probability distribution for the latter three parameters, and calculated PDFs for each of $\alpha$, $\delta$, and $T_1$ by marginalizing over the other two quantities.  We found PDFs for $\rho$ corresponding to each pulse height by calculating a histogram of all fit values, weighted by the output probability density at each corresponding grid point. PDFs for all fit geometry parameters are shown in Figure~\ref{fig:1756_geometry_pdf}.  We performed the above fit separately for each possible value of inclination, which is currently equally likely to be $\incl\degrees$ or $\inclflip\degrees$.  We find a consistent geometry in both cases; in the case of $\alpha$, the resulting distributions are mirrored, as one might expect.  A summary of our findings is included in Table~\ref{tab:1756_params}.  

\section{The evolution of the PSR~J1756$-$2251 system}
\label{sec:evol}

\begin{figure*}[tp]
  \begin{center}
    \includegraphics[width=\textwidth]{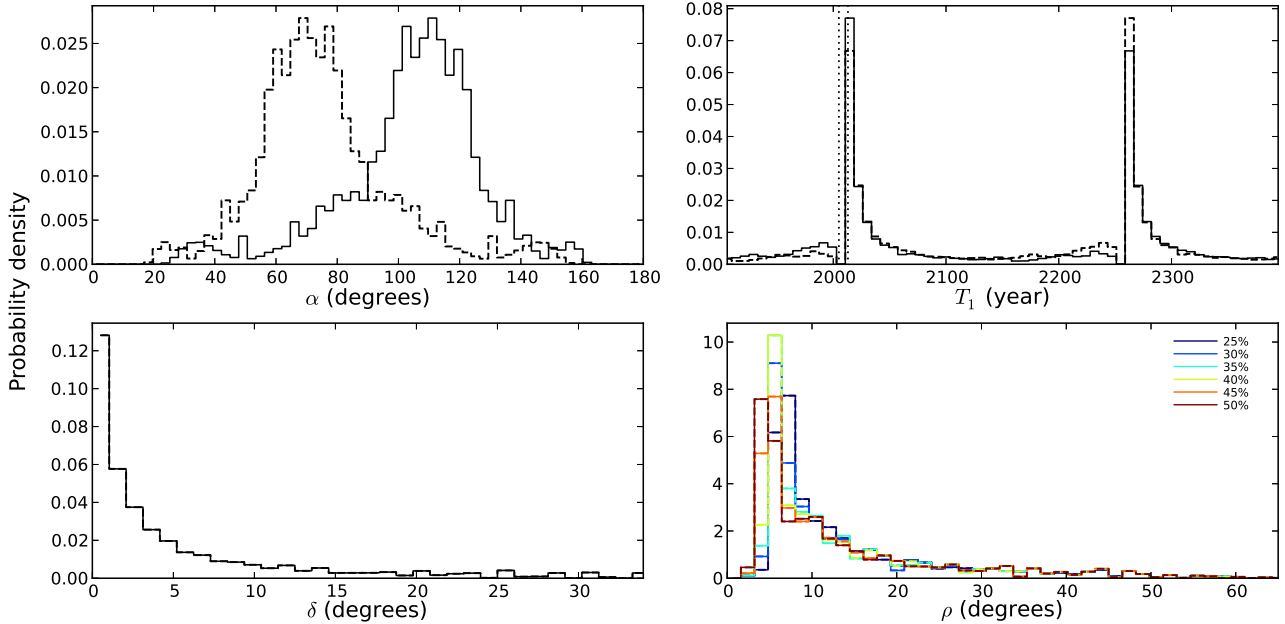}
    \caption{Results from geometry fit of PSR~J1756$-$2251 profile widths over time. Clockwise from top left are PDFs found for the angle $\alpha$ between the rotation and magnetic axes of the pulsar, the epoch of zero precession phase $T_1$, the misalignment angle $\delta$ between the pulsar spin axis and the total angular momentum of the system, and the half-opening angle $\rho$ of the portion of the pulsar beam that correspond to the pulse heights used in the fit, and are labelled in that plot.  Solid lines trace PDFs found using an inclination $i=68.0\degrees$, and dashed lines represent the case of $i=112.0\degrees$.  Dotted vertical lines plotted over the $T_1$ PDFs denote the time span of our data set.\label{fig:1756_geometry_pdf}} 
  \end{center}
\end{figure*}

From a binary evolutionary standpoint, the principal parameter of interest found from the geometry fit described in \S\ref{sec:1756_geom} is the misalignment angle $\delta$, for which we find 68\%, 95\%, and 99\% upper limits of $\deltaonesig\degrees$, $\deltatwosig\degrees$, and $\deltathreesig\degrees$, respectively.  While this is not as constraining as the $\delta$ upper limit found by \citep{fsk+13} for \psr{0737}A, it is  consistent with an alignment of the pulsar spin and total system angular momentum vectors. At the $95\%$ level, our measurement of $\delta$ is still consistent with the corresponding values for \psr{1534} \citep[$27\degrees \pm 3\degrees$;][]{fst14} and \psr{1913} \citep[$21.1\degrees \pm 0.3\degrees$;][]{wt02}.  Taken together with our timing measurements, our findings suggest that the \psr{1756} system may have proceeded through a similar evolutionary history to \psr{0737}A/B, and a different one from PSRs~B1534+12 and B1913+16.  Our findings are also consistent with the observed long-term stability of the \psr{1756} pulse profile.  As we see in Figure~\ref{fig:1756_geometry_pdf}, the $T_1$ peak PDF values 
($\sim 2004$ and $\sim 2262$ for $i = \incl\degrees$ and $\inclflip\degrees$, respectively)
occur within the time spanned by our observations. We suspect this may be the result of sparse sampling of the full precession period, and is not a reliable measurement of the epoch of zero precession phase.  Additionally, if $\delta$ is in fact nearly zero, our measurement of $T_1$ has limited meaning, as it would then be difficult to define precession phase at any epoch.  

The properties of \psr{1756} give further evidence of the existence of differing modes of DNS formation and evolution. \psr{1534} and \psr{1913}, for example, have massive companions, large eccentricities, and high transverse velocities, which indicate a high-mass loss, asymmetric supernova from a massive progenitor that imparted a significant natal kick to the system \citep[e.g.,][]{wkk00}. The \psr{1756} binary system, on the other hand, more closely resembles the double pulsar \psr{0737}A/B; along with the small misalignment angle, its low second-formed NS mass and relatively small eccentricity suggest that, as with \psr{0737}B, the NS companion to \psr{1756} may be the remnant of a low mass-loss, relatively symmetric supernova event \citep{vdh04,vdh07,wwk10,fsk+13}.  Candidates for this include the ECS and ultra-stripped He core scenarios, as discussed in \S\ref{sec:intro}.

In addition, the spin periods of \psr{1756} and \psr{0737}A are a factor of $\sim 2$ lower than those of \psr{1913} and \psr{1534}, suggesting different amounts and durations of mass transfer in spinning up the pulsars to their current rotational speeds.  Based on the observed correlation between spin period and eccentricity in most known DNS systems \citep{mlc+05,fkl+05}, it has been suggested by, e.g., \citet{dpp05} that systems which have experienced a small amount of mass loss during the second supernova are those which had lower-mass helium stars prior to that event and thus a longer timescale within which mass transfer could occur.  This, they argue, could explain the shorter spin periods as well as the low eccentricities in these systems.

One major difference between \psr{1756} and \psr{0737}A/B is the factor of 3 longer orbital period in the former system.  If the second supernova in the \psr{1756} system was indeed relatively symmetric with little mass loss, there would likely be little change in the orbital period in the resulting double neutron star binary, which was likely set by the evolution of this system prior to the second supernova under this scheme \citep[e.g.,][]{dp03b,ibk+03}.

As discussed in the introduction, a DNS system that remains bound after a low-kick, symmetric SN event is expected to have a relatively low space velocity.  The only published analysis on the natal kick velocity of \psr{1756} is by \citet{wlh06}.  The results they obtain are not very constraining for this system, since they rely only on the orbital parameters for these system, and do not include kinematic information when deriving the kick velocities.  In this work, we have measured a low proper motion in the right ascension direction, $\mu_{\alpha} = \pmraval\,$mas yr$^{-1}$.  We use our bias-corrected distance measurement (see \S\ref{sec:distance}) to calculate the velocity in the direction of right ascension, which we find to be $v_{\alpha} \sim \velra$\,km\,s$^{-1}$.  This is the same order of magnitude as the transverse velocity of the \psr{0737}A/B system, $v_{\mathrm{tr}}\sim 10$\,km\,s$^{-1}$ \citep{ksm+06}.  Proper motion in declination has been difficult to measure, since this pulsar is located very near to the ecliptic plane, making detection of proper motion in this direction a challenging task---only recently has the uncertainty approached the value quoted from timing measurements, giving $\mu_{\delta} = \pmdecval\,$mas yr$^{-1}$.  When combined with $\mu_{\alpha}$, this corresponds to a total tangential space velocity\footnote{The tangential velocities quoted in fact represent upper limits, due to the unknown contribution of differential Galactic rotation to the pulsar velocity. However, we expect the observed velocity to reflect little change to the pulsar's peculiar velocity, given the proximity of the pulsar to Earth (see Table~\ref{tab:1756_params} and \S\ref{sec:distance} for discussion and determination of distance), and assuming a flat Galactic rotation curve in the in the Solar neighbourhood.}
$v_t = \veltot$\,km\,s$^{-1}$.  Although consistent with a low value, this measurement is far from constraining; we thus quote both $\mu_\delta$ and $\mu$ as upper limits in Table~\ref{tab:1756_params}.  It is furthermore argued by \citet{kvw08} that a small transverse velocity does not necessarily imply a small velocity in the radial direction, which is very difficult to measure.  Still, if the proper motion in declination of \psr{1756} is also small, this would present another tantalizing clue that perhaps the \psr{1756} system may also have experienced a relatively small natal kick from the second supernova, as was likely the case for the double pulsar.  Further observations will thus help to resolve this issue.

The low mass loss and weak kick suffered by a star proceeding through a symmetric SN \citep{plp+04} might suggest a relatively high survival rate of DNS systems for which this is the formation mechanism for the second SN.  It may thus be the case that DNS systems that have experienced this type of SN are as common, or more common, than those formed in the aftermath of traditional ICCS events.  The binary systems that undergo the latter typically suffer a relatively large amount of mass loss and a large kick, and are thus less likely to remain bound.  

Although symmetric SN events are expected to leave behind more DNS systems intact, we have not found a greater proportion of these systems until recently.  This is likely due to a combination of several possible factors.  Firstly, survey selection effects can make some of these systems difficult to uncover.  For example, those like the double pulsar would have enhanced pulse smearing due to their small orbital period and thus large acceleration \citep[see, e.g.,][]{jk91,blw13}; this is less of a problem for systems like \psr{1756}, as well as PSRs~B1534+12 and B1913+16, which have larger orbits. 

In addition, traditional ICCS may occur more frequently than ECS or ultra-stripped helium core collapse events. It is conceivable that a system containing a low-enough mass star to eventually undergo electron capture is more likely to become unbound in the initial supernova event, resulting in fewer candidate NS-MS star systems that might otherwise evolve into systems like \psr{1756} and \psr{0737}A/B.  Although much analysis has been done in this area \citep[see, e.g.,][for studies addressing systems with low eccentricities and/or low-velocity kicks]{cb05,dpp05,ikb06,wakb08}, more work in population synthesis and binary evolutionary modeling will clearly be needed to help to address these possibilities, and could give robust estimations of relative numbers of each type of system expected to be observed.

\section{Conclusions}
\label{sec:conclusions}
We have described and presented our studies of the pulsar \psr{1756} and its host DNS system.  Through timing analysis, we have found precise measurements for the pulsar and companion NS masses, and a significantly smaller distance to the pulsar than that predicted by its measured DM value.  We have also measured a low proper motion in the right ascension direction, and although the proper motion in declination remains relatively unconstrained, this provides possible evidence for a low tangential velocity for this pulsar.  This hints at a small natal kick from the supernova that left behind the companion star.  By modeling the long-term profile shape, we have constrained the misalignment angle between the axis of rotation of the pulsar and the total angular momentum of the binary system, and find it to be consistent with the alignment of these two vectors.  Although the constraints at higher confidence levels are not yet as tight as for the double pulsar system, the perceived lack of secular changes in profile width supports a spin-orbit alignment for this system.  Taken together with the mass, eccentricity, and proper motion found through timing, this suggests a evolution for the \psr{1756} binary system that closely resembles that of \psr{0737}A/B, possibly involving the formation of the second NS via an ECS or core collapse of an ultra-stripped He core.  

Understanding the evolution of DNS systems, and the relative numbers which undergo this type of symmetric SN compared to those like PSRs~B1534+12 and B1913+16, which were born out of the more violent ICCS events, is crucial for accurately estimating the expected yields from pulsar search surveys, and more generally, in performing population synthesis calculations.  This is especially important for predicting expected source counts for the Advanced LIGO/VIRGO experiments \citep{aaa+13}, which are particularly sensitive to coalescing DNS systems.

\section*{Acknowledgements}
The authors wish to thank Caltech, Swinburne University, and NRL pulsar groups for use of the CGSR2 cluster at Green Bank, and to W.~van Straten for his help with the \texttt{psrchive} software. Thanks as well to K.~Gourgouliatos, R.~Lynch, D.~Tsang, and J.~van Leeuwen for several helpful discussions. Finally, we thank the referee, J.~Weisberg, for his very insightful comments and suggestions. Pulsar research at UBC is supported by an NSERC Discovery Grant.  CGB acknowledges support from ERC Advanced Grant ``LEAP'' (227947, PI: Michael Kramer). The Parkes radio telescope is part of the Australia Telescope which is funded by the Commonwealth of Australia for operation as a National Facility managed by CSIRO. The National Radio Astronomy Observatory is a facility of the U.S. National Science Foundation operated under cooperative agreement by Associated Universities, Inc.  GASP was partially funded by an NSERC RTI-1 grant to IHS.  The Nan\c{c}ay radio telescope is part of the Paris Observatory, associated with the Centre National de la Recherche Scientifique (CNRS), and partially supported by the R\'{e}gion Centre in France.  The Westerbork Synthesis Radio Telescope is operated by ASTRON (Netherlands Institute for Radio Astronomy), with support from the Netherlands Foundation for Scientific Research (NWO).  The Lovell Telescope is owned and operated by the University of Manchester as part of the Jodrell Bank Centre for Astrophysics, with support from the Science and Technology Facilities Council of the United Kingdom.



\label{lastpage}

\end{document}